\documentclass[manuscript]{aastex}

\begin{document}
\pagenumbering{arabic}

\title{THE LIGHT CURVE OF S ANDROMEDAE}

\author{Sidney van den Bergh}
\affil{Dominion Astrophysical Observatory, Herzberg Institute of Astrophysics, National Research Council of Canada, 5071 West Saanich Road, Victoria, BC, V9E 2E7, Canada}
\email{sidney.vandenbergh@nrc.ca}

\begin{abstract}

Historical observations of S And, in combination with the color versus rate-of-decline relationship for well observed SN 1991bg-like supernovae, are used to estimate a rate of decline $\Delta m_{15}$ (B) = 2.21 and an intrinsic color at maximum $[B(0) - V(0)]_{o}$ = 1.32 for SN 1885A.

\end{abstract}

\keywords{supernovae:individual (S Andromedae)}

\section{INTRODUCTION}

S Andromedae (1885) was probably the nearest extragalactic supernova of Type Ia to have occurred in historical times. A definitive review of the observational material that is available for this object has been given by de Vaucoulers \& Corwin (1985). On the basis of its visual light curve, and descriptions of its spectrum, these authors conclude that this object was probably an unusually fast supernova of Type I [modern Type Ia]. This conclusion is consistent with the detection of Fe I absorption at the position of SN1885A by Fesen, Hamilton \& Saken (1989). An assignment to modern Type Ia, which comprises stars with progenitors having ages $\geq$ 1 Gyr, is also supported by the fact that SN1885A occurred only $16''$ from the nucleus of M31, i.e. in a region that exhibits no traces of recent star formation. The pronounced orange color of S And near maximum light suggests that the supernova of 1885 might have been a member of the sub-luminous SN1991bg subclass (Phillips 1993) of SNe Ia. On the other hand van den Bergh (1994) tentatively concluded that the observed colors of S And, during the first weeks after maximum, were intermediate between those of normal SNe Ia and that of the sub-luminous object 1991bg. Since more data on such sub-luminous supernovae have recently become available (Phillips {\it et al.} 1999, Garnavich {\it et al.} 2001) it appears opportune to reexamine this question at the present time.

\section{LIGHT CURVE OF S ANDROMEDAE}

>From a detailed analysis of over 500 visual observations, that were subsequently tied to nearby modern photoelectric comparison sequences, de Vaucouleurs \& Corwin (1985) found that the equation

\hspace*{4cm} $V(t) = 5.85 + 1.65 [log(t - t _o)]^2$ \hspace*{5cm} (1)

\noindent gives an excellent representation of the visual light curve of S And over the range $6 < V < 14$. Unfortunately little information is available on the color of S Andromedae near maximum light. 
Only for the period 1885 Sept. 4-8 do de Vaucouleurs \& Corwin give accurate color information. An attempt will be made to use a recent compilation of information (Garnavich {\it et al.} 2001) on other fast red SNeIa to derive (1) the  color of S And at maximum light, and (2) the rate of decline $\Delta m_{15}$ (B) of S Andromedae in blue light.

>From Equ. (1) it is found that S And had V = 8.13 at 15 days past maximum, i.e. on 1885 Sept. 5.5 $\pm$ 1. According to the analysis by de Vaucouleurs \& Corwin (1985) S And had a color B-V = 1.31 on this date. It follows that B(15) = 9.44. Unfortunately the data analyzed by de Vaucouleurs {\it et al.} do not directly give B(0), the blue magnitude of S Andromedae at maximum light. However, it is known that (1) S And was unexpectedly red in the weeks after maximum, and (2) that this object had an unusually high rate of luminosity decline. These factors suggest that S And belonged to the SN 1991bg class of SNeIa. 
All available photometric data on nine well-observed supernovae that appear to belong to this class have recently been discussed by Garnavich {\it et al.} (2001). From such data these authors find a well-defined relationship between color at maximum light and rate of decline for sub-luminous supernovae resembling SN 1991bg. Garnavich {\it et al.} find that nine well- observed objects of this type with  $1.7 < \Delta m_{15}$ (B) $<$ 2.0 follow the relation 

\hspace*{4cm} $[B(0) - V(0)]_o = 2.38 \Delta m_{15} - 3.95.$ \hspace*{5cm} (2)\\

\noindent This may be recast in the form

\hspace*{2cm} $B(0)_o$ = $V(0)_o$ + 2.38 $[ B(15)_o$ - $B(0)_o$ ] - 3.95. \hspace*{5cm} (3)

\noindent From B(15) = 9.44 and A$_B$ = 0.25 mag (van den Bergh 2000) one has $B(15)_o = 9.19$. Furthermore from V(0) = 5.85 (de Vaucoulers \& Corwin 1985) and $A _V$ = 0.19 mag one finds $V(0)_o = 5.66$. Substituting these values into Equ. 3 and solving for $B(0)_o$  one finds that $B(0)_o$  = 6.98. In conjunction with the value $B(15)_o$  = 9.19 that was obtained above this yields $\Delta m_{15}$ (B) = 2.21 mag. This value is larger than that of any of the nine SN 1991bg -like SNeIa for which accurate photoelectric photometry is presently available. However, de Vaucouleurs \& Corwin point out that an even larger value $\Delta m _{15}$ (B) was observed for SN 1939b in the E5 galaxy NGC 4621(=M59). From $V(0)_o$= 5.66 and $B(0)_o$ = 6.98 one finds $[B(0) - V(0)]_o$ = 1.32 for S Andromedae. This value is close to the (uncertain) color index $(B-V)_o$ =  1.25 at 15 days past maximum that was obtained by de Vaucouleurs \& Corwin. In other words S And appears to have exhibited little color change during the first two weeks after maximum. In this respect S And seems to differ from the prototypical object 1991bg (Leibundgut {\it et al.} 1993) which had B-V = 0.85 at maximum, but then reddened rapidly to B-V $\sim$ 1.25 by five or six days after  maximum.

\section{CONCLUSIONS}

By combining the observed visual light curve of S And, and its color in early September of 1885, with the color versus rate-of-decline relation of well-observed sub-luminous SNeIa one may estimate the color at maximum and the rate of decline of S And in blue light.  Adopting $A_V$ = 0.19 mag , $A_B$ = 0.25 mag and a distance modulus $(m-M)_o$ = 24.4 $\pm$ 0.1 (van den Bergh 2000) for M31, in conjunction with the hypothesis that S And was a sub-luminous supernova of the 1991bg class, one then finds that, at maximum light:

$V(0)_o$  = 5.66 $\pm$ 0.5   

$M(0)_V$ = -18.74 $\pm$ 0.5  

$B(0)_o$ =  6.98  

$M(0)_B$  = -17.42, and hence  

$[B(0) -V(0)]_o$  = 1.32  

\noindent The initial rate of decline in blue light is found to be $\Delta m_{15}$ (B) = 2.21 $\pm$ 0.1 mag. The errors of the magnitudes that are quoted above  are not well determined by the observational data obtained in 1885. It is noted in passing that the values of $M(0)_V$ and $M(0)_B$ obtained here for S Andromedae lie significantly above the trend lines in $M(0)_V$ versus $\Delta m_{15}$ (B), and $M(0)_B$ versus $\Delta m_{15}$ (B) that were plotted by Garnavich {\it et al.} (2001) in their Fig.16. By the same token S And lies significantly above the $M(0)_B$ versus $[B(0) - V(0)]_o$ , and the $M(0)_V$ versus $[B(0) - V(0)]_o$ relations that Garnavich {\it et al.} show in their Figure 17. To fit the trend lines shown in their figure our V(0) value for S And would have to be too bright by $\sim$2.1 mag, or our B-V color at maximum would have to be too red by $\sim$1.3 mag. Such large values appear inconsistent with the historical data compiled by de Vaucouleurs \& Corwin (1985). Taken at face value these results  would seem to indicate that there may be significant intrinsic luminosity and/or color dispersion among those SNe Ia having the  fastest rates of luminosity decline. Alternatively, but less probable, S And might differ from the Sne Ia studied by Garnavich {\it  et al.} 

It is a pleasure to thank David Branch, Mark Phillips and Nick Suntzeff for helpful exchanges of e-mail about supernovae of Type Ia.

\end{document}